\documentclass[11pt,twocolumn]{emulateapj_Astroph}

\tightenlines
\usepackage{epstopdf}
\usepackage{ulem} 
\usepackage{color} 
\usepackage{lineno}

\def\lsim{\;\raise0.3ex\hbox{$<$\kern-0.75em\raise-1.1ex\hbox{$\sim$}}\;}
\def\gsim{\;\raise0.3ex\hbox{$>$\kern-0.75em\raise-1.1ex\hbox{$\sim$}}\;}

\definecolor{purple}{RGB}{200,100,255} 

\newcommand{\alfL}{\alpha_L}
\newcommand{\alfH}{\alpha_H}
\newcommand{\alfP}{\alpha(p)}

\newcommand{\gamray}{$\gamma$-ray}

\newcommand{\GRBs}{$\gamma$-ray bursts}

\newcommand{\UM}{unmodified}

\newcommand{\GBeta}{\gamma(x)\beta(x)}
\newcommand{\GBetaZ}{\gamma_0\beta_0}

\newcommand{\pFEB}{p_\mathrm{FEB}}

\newcommand{\LfebUpS}{L_\mathrm{UpS}}

\newcommand{\delMax}{\delta\theta_\mathrm{max}}

\newcommand{\delAng}{\delta\theta}

\newcommand{\PAD}{pitch-angle distribution}
\newcommand{\PADs}{pitch-angle distributions}
\newcommand{\PAS}{pitch-angle scattering}

\newcommand{\gamZ}{\gamma_0}

\newcommand{\Nonrel}{Non\-rel\-a\-tiv\-is\-tic}

\newcommand{\ultrarel}{ul\-tra-rel\-a\-tiv\-is\-tic}

\newcommand{\Lor}{Lorentz}

\newcommand{\Lmfp}{\lambda_\mathrm{mfp}}

\newcommand{\pion}{pion-decay}

\newcommand{\aaT}{{\bf A}}
\newcommand{\bBB}{{\bf B}}
\newcommand{\cCC}{{\bf C}}
\newcommand{\dDD}{{\bf D}}
\newcommand{\eEE}{{\bf E}}
\newcommand{\fFF}{{\bf F}}
\newcommand{\gGG}{{\bf G}}
\newcommand{\hHH}{{\bf H}}
\newcommand{\iII}{{\bf I}}
\newcommand{\jJJ}{{\bf J}}
\newcommand{\kKK}{{\bf K}}
\newcommand{\lLL}{{\bf L}}
\newcommand{\mMM}{{\bf M}}
\newcommand{\nNN}{{\bf N}}
\newcommand{\oOO}{{\bf O}}
\newcommand{\degg}{^\circ}

\newcommand{\Ldiff}{L_\mathrm{diff}}

\newcommand{\EffDSA}{{\cal E}_\mathrm{DSA}}

\newcommand{\xx}[1]{\!\times\!10^{#1}}

\newcommand{\Lfeb}{L_\mathrm{FEB}}
\newcommand{\LfebP}{L_\mathrm{FEB}^\prime}
\newcommand{\pcc}{cm$^{-3}$}
\newcommand{\kmps}{km s$^{-1}$}

\newcommand{\SA}{semi-analytic}

\newcommand{\RH}{Rankine-Hugoniot}

\newcommand{\mfp}{mean free path}

\newcommand{\PIC}{Particle-in-cell}
\newcommand{\pic}{particle-in-cell}

\newcommand{\pD}{p_d}
\newcommand{\Pd}{p_d}
\newcommand{\PdP}{p_d^\prime}
\newcommand{\rg}{r_g}
\newcommand{\rgz}{r_{g0}}

\newcommand{\gyrotime}{\tau_g}
\newcommand{\deltime}{\delta t}
\newcommand{\etamfp}{\eta_\mathrm{mfp}}

\newcommand{\Rtot}{R_\mathrm{tot}}

\newcommand{\nonrel}{non-relativistic}

\newcommand{\rel}{relativistic}

\newcommand{\pmax}{p_\mathrm{max}}

\newcommand{\mc}{Monte Carlo}
\newcommand{\MC}{Monte Carlo}

\newcommand{\Facc}{Fermi acceleration}
\newcommand{\FoFSA}{first-order Fermi shock acceleration}

\newcommand{\TP}{test-particle}
\newcommand{\NL}{nonlinear}
\newcommand{\SCly}{self-consistently}
\newcommand{\SC}{self-consistent}

\renewcommand{\baselinestretch}{1}
\def\aap{Astronomy and Astrophysics}

\def\apjl{The Astrophysical Journal Letters}

\newcount\listnorom
\newcommand\listromanDE{\global\advance \listnorom by 1
{\lowercase\expandafter{(\romannumeral\listnorom)}\ }}

\newcount\listnumber
\newcommand\listDE{\global\advance \listnumber by 1
{\lowercase\expandafter{(\number\listnumber)}\ }}
\newcommand\newlistDE{\listnumber=0}
\newcount\Inum
\newcount\IInum
\newcount\IIInum
\newcount\IVnum
\def\I{\global\multiply\IInum by 0 \global\multiply\IIInum by 0
            \global\multiply\IVnum by 0 \global\advance \Inum by 1
            {\the\Inum. }}
\def\II{\global\multiply\IIInum by 0\global\multiply\IVnum by 0
       \global\advance \IInum by 1 {\the\Inum.\the\IInum. }}
\def\III{\global\multiply\IVnum by 0\global\advance \IIInum by 1
            {\the\Inum.\the\IInum.\the\IIInum. }}
\def\IV{\global\advance \IVnum by 1
            {\the\IVnum. }}
\shorttitle{Particle Scattering in Relativistic Shocks}
\shortauthors{Ellison, Warren \& Bykov}

\begin{document}

\title{Particle spectra and efficiency in nonlinear relativistic shock acceleration: survey of scattering models} 

\vskip24pt

\author{Donald C. Ellison,\altaffilmark{1}
Donald C. Warren\altaffilmark{2} and
Andrei M. Bykov\altaffilmark{3,4,5}}

\altaffiltext{1}{Physics Department, North Carolina State
University, Box 8202, Raleigh, NC 27695, U.S.A.;
don\_ellison@ncsu.edu}

\altaffiltext{2}{Astrophysical Big Bang Laboratory, RIKEN, Saitama 351-0198, Japan; donald.warren@riken.jp}

\altaffiltext{3}{Ioffe Institute for Physics and Technology, 194021
St. Petersburg, Russia; ambykov1@gmail.com}

\altaffiltext{4}{International Space Science Institute, Bern, Switzerland}

\altaffiltext{5}{Saint-Petersburg State Polytechnical University,Saint-Petersburg, Russia}


\begin{abstract}
We include a general form for the scattering \mfp, $\Lmfp(p)$, in a \NL\ \mc\ model of \rel\ shock formation and \Facc. \PIC\ (PIC) simulations, as well as analytic work, suggest that \rel\ shocks tend to produce short-scale,  
self-generated magnetic turbulence that leads to a scattering \mfp\ with a stronger momentum dependence than the $\Lmfp \propto p$ dependence for Bohm diffusion. 
In unmagnetized shocks, this turbulence is strong enough to dominate the background magnetic field so the shock can be treated as parallel regardless of the initial magnetic field orientation, making application to  \GRBs\ (GRBs), pulsar winds, Type Ibc supernovae, and extra-galactic radio sources more straightforward and realistic.
In addition to changing the scale of the shock precursor, we show that, when \NL\ effects from efficient \Facc\ are taken into account, the momentum dependence of $\Lmfp(p)$ has an important influence on the efficiency of cosmic-ray production as well as the accelerated particle spectral shape.
These effects are absent in \nonrel\ shocks and do not appear in \rel\ shock models unless \NL\ effects are \SCly\ described. 
We show, for limited examples, how the changes in \Facc\ translate to changes in the intensity and spectral shape of \gamray\ emission from proton-proton interactions and \pion\ radiation. 
\end{abstract}

\section{Introduction}
Fermi shock acceleration at \rel\ shocks may  be an important 
cosmic-ray (CR) production mechanism in a number of sites including \GRBs\ (GRBs), pulsar winds, type Ibc supernovae, and extra-galactic radio sources \citep[see][for a review]{BykovEtal2012}. 
Central to \Facc, as well as all other aspects of \rel\ shock physics, is the production of the magnetic turbulence  responsible for collisionless wave-particle interactions. 
Because of this, a great deal of work has been devoted studying 
plasma instabilities that may be important near \rel\ shocks
\citep[see][and references 
therein]{LP2010,RKW2011,PPL2013,SSA2013,LPGP2014,RevilleBell2014}.

Despite this intense effort, the issue of particle 
transport in \rel\ plasmas is by no means settled.
In principle, \pic\ (PIC) simulations 
can solve the full problem of shock formation, wave generation, and particle acceleration \SCly; and a great deal of work has been done with these techniques
\citep[e.g.,][]{KT2008,Spitkovsky2008,Nishikawa2009,SSA2013}.
These simulations typically show the Weibel instability producing magnetic turbulence on skin-depth scales in the subshock vicinity  
and this self-generated turbulence can then lead to superthermal particle acceleration
\citep[e.g.,][]{KeshetEtal2009,Haugbolle2011,SironiSpit2011}.

The short-scale Weibel instability is critical in the initial shock formation process seen in PIC simulations, but \citet{LPGP2014}, for example, have shown that other longer-scale instabilities may develop in the shock precursor that dominate Weibel and modify the precursor structure in important ways.
There are also indications from GRB afterglow observations that rapid 
\Facc\ and Bohm diffusion may occur in \ultrarel\ shocks 
\citep[][]{SagiNakar2012}.
While PIC simulations are the only way to obtain fully \SC\ solutions, the computational limitations of PIC simulations 
\citep[see][]{VBE2008} may be such that important \NL\ (NL) effects of \Facc\ on long-wavelength instabilities  and shock precursor
structure are currently beyond the  reach of this technique.

In this paper we investigate the kinematics of \FoFSA\ in \rel\ shocks using a  parameterized, momentum-dependent, scattering mean free path, $\Lmfp(p)$, where $p$ is the local frame particle momentum. 
In particular, we examine the difference between a Bohm momentum dependence, i.e., $\Lmfp \propto p$, and $\Lmfp \propto p^{\alfP}$, where $\alpha=2$ is expected from scattering off small-scale turbulence such as that generated by the Weibel instability 
\citep[e.g.,][]{Jokipii72,PPL2011,PPL2013}. 
While our parameterization cannot address fundamental issues concerning wave generation and shock formation, as done by PIC simulations or some \SA\ models \citep[e.g.,][]{marcowith06,PLM2009,RKW2011,Casse2013,RevilleBell2014,Schlickeiser2015}, we can model large spatial and momentum dynamic ranges and show how efficient \Facc\ varies with the momentum dependence of $\Lmfp(p)$ when the backreaction of CRs on the shock precursor  is included \SCly.

The \mc\ simulation we employ has been described in detail in 
\citet{EWB2013,WarrenEllison2015}, and references therein 
\citep[see][for an early non-PIC 
discussion of NL \Facc\ in \rel\ shocks]{ED2002}. 
With this technique, the complex plasma physics of shock formation, magnetic turbulence generation, 
and particle acceleration \citep[e.g.,][]{SSA2013} is contained in the assumptions made to describe \PAS\ and in our parameter $\alfP$.
We emphasize that the \mc\ simulation models particle transport not diffusion; an important distinction in \rel\ shocks. No diffusion approximation is required and we can accurately follow particles as they scatter in plasmas moving at \rel\ speeds with large velocity gradients.

Our main result is that, apart from a large change in length scale and the subsequent drop in the maximum accelerated particle energy for finite systems, 
the momentum dependence of $\Lmfp(p)$ influences the shock structure and CR production in a purely \rel\ fashion. We describe effects that do not occur in \nonrel\ shocks, or in \rel\ shocks where the backreaction of CRs on the shock structure is ignored.
We find, for a given maximum CR momentum, $\pmax$, set by a finite shock size, that a strong momentum dependence for $\Lmfp$ in NL shocks can produce a significant {\it increase} in the \Facc\ efficiency compared to that predicted with Bohm diffusion. This effect depends on the detailed momentum dependence of $\alfP$.

As expected, any variation in the particle acceleration efficiency translates to a variation in the photon emissivity. We show, for some test parameters, that \NL\ effects can produce a  factor of three enhancement, and a noticeable change in spectral shape, in the $10-100$\,GeV \pion\ emissivity between Bohm diffusion and $\alfP\sim 2$.

\section{Details of Model}
In our steady-state \mc\ model, particles, regardless of their momentum or position relative to the subshock, are assumed to interact with a turbulent background magnetic field so their pitch-angle-scattering  mean free path is
\begin{eqnarray} \label{eq:mfp}
\Lmfp(p) &=& 
\etamfp \cdot \rg(\pD) \cdot (p/\pD)^{\alfL} \quad \mathrm{for} 
\quad p < \pD
\nonumber \\
&=& 
\etamfp \cdot \rg(\pD) \cdot (p/\pD)^{\alfH} \quad \mathrm{for} 
\quad p \ge \pD
\ .
\end{eqnarray}
Here $p$ is the particle momentum in the local frame, $\alfL \le 1$, $\alfH \ge 1$, $\pD$ is a dividing momentum between the 
low- and high-momentum ranges, $\rg(\pD)=\pD c/(eB_0)$ is the gyroradius for a proton with local-frame momentum $\pD$ in the background field $B_0$, and $\etamfp \ge 1$ is a parameter that determines the strength of scattering.\footnote{For simplicity, we replace a general $\alfP$ with the broken power law form shown in equation~(\ref{eq:mfp}). More complicated forms for $\Lmfp$ can easily be used to model \SA\ and/or PIC results where available.} 
The Bohm limit is described by $\etamfp = \alfL = \alfH = 1$, and the conditions $\alfL \le 1$ and $\alfH \ge 1$ ensure that $\Lmfp \ge \etamfp \rg$ for all $p$.\footnote{For simplicity we 
sometimes refer to ``Bohm diffusion" when we take $\etamfp=\alfL=\alfH=1$. As for any choice of parameters in equation~(\ref{eq:mfp}), particle trajectories are calculated in the \mc\ code without making a diffusion approximation.}
In the simple geometry of our plane-parallel, steady-state model, results scale simply with $\etamfp$ and we set $\etamfp=1$ in all that follows.
Discussions of the influence of $\etamfp$ in 
unmodified, oblique \rel\ shocks can be found in 
\citet{Ostrowski1991,ED2004,DoubleEtal2004,SummerlinBaring2012} 
\citep[see][for a discussion of $\etamfp$ in GRB afterglows]{SagiNakar2012}.

In the  prescription given by equation~(\ref{eq:mfp}), the background field $B_0$ (arbitrarily taken here to be $10^{-4}$\,G) is assumed to be parallel to the shock normal and only sets the scale of the shock through $\rg(\pD)$. 
Recent PIC 
simulations \citep[i.e.,][]{SSA2013} have shown that unmagnetized, \rel\ shocks, regardless of their initial magnetic field orientation, develop strong, self-generated magnetic turbulence which quickly dominates the background field. The unmagnetized condition should apply to external afterglow shocks in GRBs, early-phase shocks in Type Ibc supernovae, and  perhaps other astrophysical sites such as radio jets and pulsar winds. In these cases, the background field can be treated as parallel as we do here.

While equation~(\ref{eq:mfp}) is a gross simplification of the 
wave-particle interactions that occur in \rel\ plasmas, it does allow the  investigation of basic particle transport  effects in \rel\ flows.
Furthermore, if \rel\ shocks accelerate CRs  efficiently to ultra-high energies, as is often assumed 
\citep[e.g.,][]{Kulkarni1999,KW2005,Globus2015}, important kinematic effects from momentum and energy conservation will occur regardless of the details of the plasma physics. These kinematic effects can be investigated with \mc\ techniques.

The only modification in the \mc\ model described 
in \citet{WarrenEllison2015} is that here we use 
equation~(\ref{eq:mfp}) instead of $\Lmfp=\rg$.
Pitch-angle scattering is modeled as follows: after a 
time $\deltime \ll \gyrotime$ ($\gyrotime$ is the gyro-period) a particle scatters isotropically and elastically in the local plasma frame through a small angle $\delAng \le \delMax$.
The maximum scattering angle in any scattering event is given by 
\citep[see][]{EJR90}
\begin{equation} \label{eq:Tmax}
\delMax = \sqrt{6 \deltime/t_c}
\ ,
\end{equation}
where $\deltime=\gyrotime/N_g$, $t_c=\Lmfp/v$ is the collision time, $v$ is the particle speed in the local frame, and 
$N_g$ is a free parameter.
In all examples here, $N_g$ is chosen large enough to 
produce fine-scattering results that do not change substantially as $N_g$ is increased further \citep[see][for \TP\ examples with values of $N_g$ producing large-angle scattering]{SummerlinBaring2012}.

All particles are injected far upstream with a thermal distribution and 
scatter  and convect into and across the subshock into the 
downstream region.\footnote{The subshock, set at $x=0$, is indistinguishable from the full shock for \UM\ and \TP\ cases. It is an abrupt transition where thermal particles receive most of their entropy boost upon crossing into the downstream region. With efficient \Facc, the shock is  smoothed on larger scales from backstreaming CRs but a relatively abrupt subshock still exists where quasi-thermal particles receive a large entropy boost. We assume the subshock is transparent, i.e., we do not attempt to describe the effects of a cross-shock potential, amplified magnetic turbulence, or other effects that may occur in the viscous subshock layer.} 
Upon interacting with the downstream plasma, a particle gains energy sufficient to allow it to scatter back across the subshock and be further accelerated. The fraction of particles that are injected, i.e., do manage to re-cross the subshock is determined stochastically and constitutes our ``thermal leakage injection" model once the  further assumption is made that the subshock is transparent.  
This injection model requires no additional parameters or assumptions once the scattering mean free path is defined in equation~(\ref{eq:mfp}). 

For our steady-state simulations, the  $\pmax$ a shock can produce is determined by an upstream free escape boundary (FEB) parameter, $\Lfeb$, where $\Lfeb$ is the distance from the subshock to the FEB. 
In our plane-parallel shock approximation, a FEB  mimics the effect of high-energy particles ``leaking" from a finite-size 
shock \citep[see][for a fuller discussion of particle escape]{Drury2011}. 
In a finite shock, high-energy particles far upstream from the subshock have a high probability of escaping since the level of 
self-generated turbulence must decrease with upstream distance from the subshock. Particle escape could as well occur downstream from the subshock, as assumed in \rel\ shock calculations by \citet{Achterberg2001} and \citet{WarrenEllison2015}, or  from the sides, as might be important in jets. 
How escape occurs will depend on the detailed geometry of the shock but all these scenarios produce a corresponding $\pmax$ with no important differences from what we describe here. An upstream FEB has been used extensively in models of CR production in young SNRs \citep[e.g.,][]{EPSBG2007,MAB2009} where there is typically a SNR age where a transition occurs between $\pmax$ being determined by the remnant age to being determined by the finite shock radius. We note that there is direct observational evidence for upstream escape at the quasi-parallel Earth bow shock 
\citep[e.g.,][]{Trattner2013}.

With $\Lfeb$ set, the shock structure is determined so momentum and energy are conserved \citep[see][for details]{EWB2013}. The injection rate adjusts consistently and the \NL\ examples we show below conserve momentum and energy to within a few percent.

We note that for unmodified (UM) shocks, where the backreaction of CRs is ignored, the model injects particles into the Fermi mechanism with an efficiency that is inconsistent with a discontinuous shock structure. With our injection model, momentum and energy can only be conserved if the backreaction of CRs on the shock structure is taken into account.

We further note that our UM examples are not \TP\ (TP) ones, although TP predictions can be easily inferred from our UM results, as described in \citet{WarrenEllison2015}. In a TP shock, injection must be weak enough so Fermi accelerated CRs are few enough so  their influence on the shock structure can be ignored. The CR spectral shapes predicted for TP shocks are only applicable if an insignificant amount of energy is put into CRs. 

Even though shock smoothing must 
occur if \Facc\ is efficient, we include UM/TP examples for several reasons. 
First, there is a universally accepted  prediction for the power law slope for \Facc\ in UM/TP \rel\ shocks that must be reproduced for a method to be valid \citep[e.g.,][]{BO98,Achterberg2001}. We obtain this result within statistical limits. 
Second, the most obvious manifestation of the momentum dependence of $\Lmfp$ is a simple scaling which can be most directly demonstrated with UM results.
Finally, most other work on \rel\ \Facc,  other than PIC simulations, has been done in the TP approximation. We contrast UM and NL results to emphasize where \NL\ effects become important.

\begin{figure}
\epsscale{0.95}
\plotone{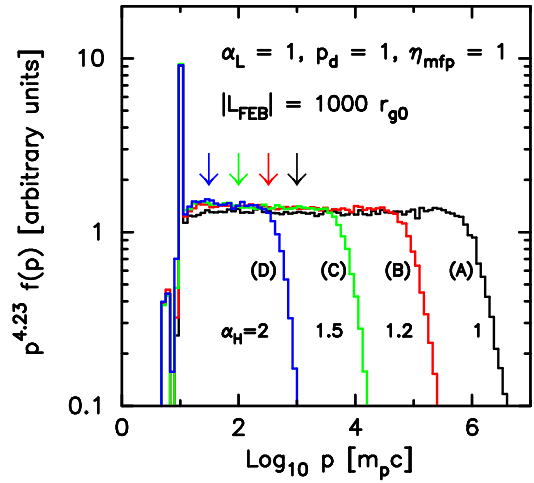}           
\caption{Phase-space distributions measured at $x=0$ in the shock rest frame for various momentum dependencies for the mean free path parameters as indicated 
(see equation~\ref{eq:mfp}). 
The letters refer to the models listed in Table~\ref{tab:param}.
In all cases, an upstream FEB at $x=-1000\,\rgz$ produced the high-momentum cutoff. Note that $p^{4.23}f(p)$ is plotted where $\sigma=4.23$ is the canonical \TP\ power-law index for \rel\ shock acceleration. 
The color-coded arrows indicate the momenta where $\Lmfp(p)=|\Lfeb|$. These momenta are shown as dots in Fig.~\ref{fig:mfp_vary}. In plane-parallel, \nonrel\ shocks, such as models \nNN\ and \oOO\ discussed below, the arrows would be at $\pmax$ regardless of $\alfH$.  
\label{fig:fp_UM}}
\end{figure}

\subsection{Particle Transport vs. Diffusion}
Equations~(\ref{eq:mfp}) and (\ref{eq:Tmax}), along with the assumptions that particles scatter isotropically and elastically in the local plasma frame, in each scattering event, fully determine the particle transport in the \mc\ simulation.
As shown in \citet{EJR90}, equation~(\ref{eq:Tmax}) ensures that, on average, a particle in a uniform flow moves a displacement $\Lmfp$ in acquiring a net deflection of $\sim 90\degg$.

This particle transport model is more general than the 
diffusion-advection equation
which is widely used in \Facc\ modeling. The \mc\ transport model accounts for an arbitrary particle anisotropy while the diffusion-advection equation accounts only for current anisotropy (i.e., the first Legendre polynomial in the particle \PAD). If the scattering rate is high enough, and the background flows are extended and uniform, than to the first approximation both models would provide the same results. However, in weak scattering and/or in highly non-uniform, \rel\ flows, the \mc\ transport treatment is more accurate.
For example, a downstream particle crossing into the upstream region can, upon interacting with the upstream flow, suffer a small deflection and immediately be overtaken by the shock. Non-diffusive particle transport of this  nature is essential for describing \Facc\ and is fully contained 
in equations~(\ref{eq:mfp}) and (\ref{eq:Tmax}).

\section{Results}
\subsection{Unmodified Shocks} \label{sec:UM}
For UM shocks, the bulk flow is discontinuous with a transition from 
shock-frame speed
$u_0$ for $x<0$ to $u_2$ for $x>0$. The shock compression ratio is defined as $\Rtot = u_0/u_2 \simeq 3.02$ for the 
$\gamZ=[1 - (u_0/c)^2]^{-1/2} = 10$ examples 
shown in Fig.~\ref{fig:fp_UM} 
\citep[see][for a detailed explanation of how  
$\Rtot$ is determined]{DoubleEtal2004}.
In Fig.~\ref{fig:fp_UM} we show particle phase-space distributions  $f(p)$ ($\times p^{4.23}$) calculated in the shock rest frame at $x=0$ for different values of $\alfH$, all with $\alfL=1$, $\pD= m_p c$, $\etamfp=1$, and an upstream FEB at $\Lfeb=-10^3\,\rgz$, where the scaling factor $\rgz \equiv \etamfp m_p u_0 c/(e B_0)$.\footnote{In the shock rest frame the plasma flows in the positive $x$-direction so an upstream FEB is at negative $x$ measured from the subshock at $x=0$. While our definition of $\rgz$ includes the scaling factor $\etamfp$ it does not include the particle \Lor\ factor. We only consider protons here and $m_p$ is the proton mass. A discussion of ion and electron acceleration is given in \citet{WarrenEllison2015}.}
These examples assume an infinite downstream region and $\Lfeb$ solely determines $\pmax$. The letters in Fig.~\ref{fig:fp_UM} indicate the models as listed in Table~\ref{tab:param}.

\begin{figure}
\epsscale{1.0}
\plotone{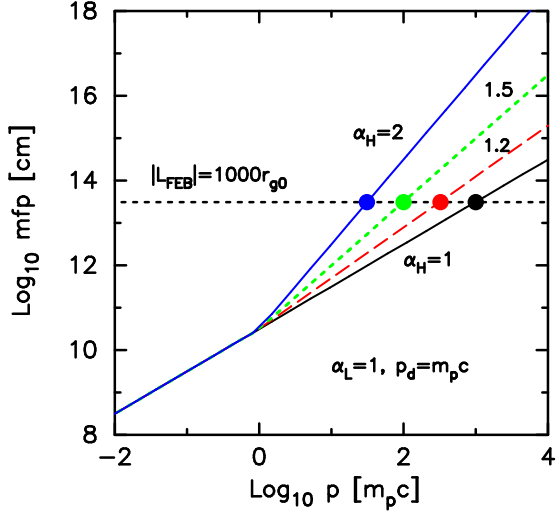}           
\caption{Mean free path for the models \aaT$-$\dDD\ shown in 
Fig.~\ref{fig:fp_UM} as indicated by $\alfH$. The upstream FEB is shown as a horizontal dotted line. The dots highlight the momentum needed in each case to produce a $\Lmfp$ equal to $|\Lfeb|$. 
\label{fig:mfp_vary}}
\end{figure}

The mean free paths (equation~\ref{eq:mfp}) for Fig.~\ref{fig:fp_UM} are shown in Fig.~\ref{fig:mfp_vary}, where the horizontal dotted line is the position of the upstream FEB, i.e., 
$\Lfeb =-10^3\,\rgz \simeq -10^{-5}$\,pc for the particular value $B_0=10^{-4}$\,G chosen for these examples. 
The main effect in going from the Bohm limit 
($\alfH=1$) to $\alfH=2$ is the decrease in $\pmax$ as the diffusion length at a given momentum increases with $\alfH$ and becomes comparable to $\Lfeb$.
Except for some 
subtle effects at momenta just above $\gamZ m_p c$, the superthermal spectral index has the canonical value $4.23$ and is independent of $\alfH$, as is the thermal leakage injection efficiency, in these UM examples. We discuss injection effects in more detail in Section~\ref{sec:Inj}.

\subsection{Precursor Transport}
Particle transport in the \rel\ shock precursor differs considerably from that in \nonrel\ shocks.
For \nonrel\ flows, the upstream diffusion length in an unmodified shock is
\begin{equation} \label{eq:Ldiff}
\Ldiff = D/u_0 = \Lmfp v/(3 u_0)
\ ,
\end{equation}
where $D=\Lmfp v/3$ is the diffusion coefficient. Setting $\Ldiff=|\Lfeb|$, defining $\LfebP = |\Lfeb|/\rgz$, 
$\PdP  =  \Pd/(m_p  c)$, and taking $v \simeq c$,  we have
\begin{equation} \label{eq:Lfeb}
\LfebP =\frac{\PdP}{3} 
\left ( \frac{c}{u_0} \right )^2
\left ( \frac{\pFEB}{\Pd} \right )^{\alfH}
\ ,
\end{equation}
where
\begin{equation}
\Lmfp = \etamfp \cdot \rg(\pD) \cdot (p/\pD)^{\alfH}
\ .
\end{equation}
Therefore, the momentum associated with the diffusion distance to the upstream FEB is
\begin{equation} \label{eq:Pfeb}
\pFEB = \PdP
\left [ \frac{3 \LfebP}{\PdP} \left(\frac{u_0}{c}\right)^2
\right ]^{1/\alfH} m_p c
\ .
\end{equation}
We include this well known result to emphasize that, while it is of fundamental importance for \nonrel\ shocks and $\pFEB \simeq \pmax$ regardless of $\alfH$ in the subset of \nonrel\ shocks where  \Facc\ is limited by an upstream FEB 
(see Section~\ref{sec:NR}), 
it does not apply for \rel\ flows. 

In Fig.~\ref{fig:fp_UM},
$\PdP=1$, $\LfebP=1000$, $u_0 \simeq c$, and we can estimate $\pmax$ for each $\alfH$ from the curves.
We find: 
$(\pFEB/\pmax)_{\alfH=1} \simeq (3000/4\xx{5}) \simeq 7.5\xx{-3}$,
$(\pFEB/\pmax)_{\alfH=1.2} \simeq (790/5\xx{4}) \simeq 0.016$,
$(\pFEB/\pmax)_{\alfH=1.5} \simeq (210/5\xx{4}) \simeq 0.05$, and
$(\pFEB/\pmax)_{\alfH=2} \simeq (55/300) \simeq 0.2$.
The fact that $\pFEB/\pmax \ll 1$, and $\pFEB/\pmax$ depends strongly on $\alfH$, shows the non-diffusive behavior of particle transport in the \rel\ shock precursor as captured by the \mc\ simulation. 
In  is noteworthy that, as the momentum dependence of $\Lmfp$ increases, the particle transport approaches the \nonrel\ 
diffusion length $D/u_0$.
It is also noteworthy that the relation between $\Lfeb$ and $\Ldiff$ will vary continuously as the shock \Lor\ factor drops as it will during the GRB afterglow phase.

\begin{figure}
\epsscale{1.0}
\plotone{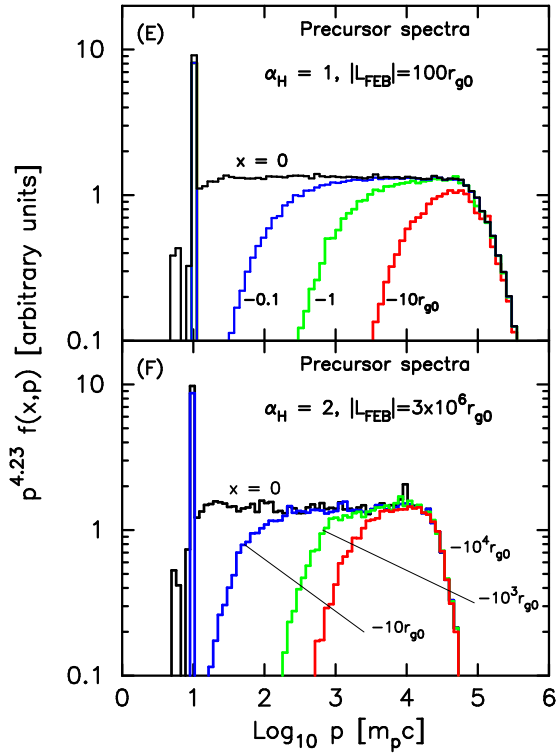}           
\caption{The top panel shows shock-frame precursor spectra for an $\alfH=1$ example  with $\Lfeb=-100\,\rgz$
(Model~\eEE). The position ahead of the subshock where $f(x,p)$ was calculated is indicated for each curve. The bottom panel shows shock-frame precursor spectra for Model~\fFF\ with $\alfH=2$, $\Pd=m_p c$, and $\Lfeb=-3\xx{6}\,\rgz$.
\label{fig:fp_VaryX}}
\end{figure}

\begin{figure}
\epsscale{0.95}
\plotone{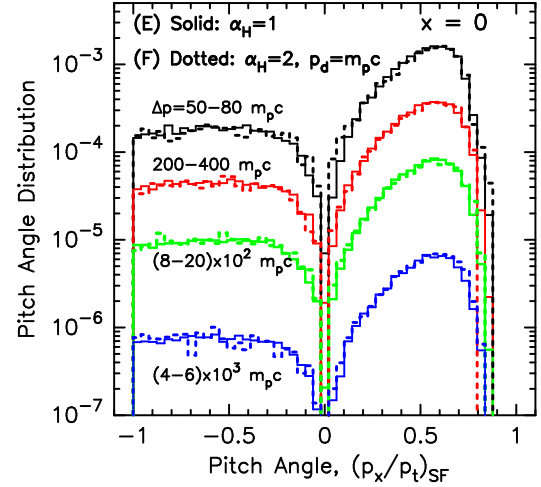}  
\caption{Pitch-angle distributions for the $\alfH=1$ 
(solid curves, Model~\eEE) and $\alfH=2$, $\Pd=m_pc$ 
(dotted curves, Model~\fFF) examples shown in 
Fig.~\ref{fig:fp_VaryX} calculated at $x=0$. The distributions are 
calculated in the shock frame for four momentum ranges as indicated. All curves are normalized to the total injected particle flux so they are absolutely normalized relative to each other. The ratio 
$(p_x/p_t)_\mathrm{SF}$ is the $x$-component of momentum divided by the 
total momentum measured in the shock-rest frame. A particle 
with  $(p_x/p_t)_\mathrm{SF}=1$ points directly downstream and, in the shock rest frame, no particle crosses $x=0$ with $(p_x/p_t)_\mathrm{SF}=0$.
\label{fig:PA_L1_L2}}
\end{figure}

\subsection{Precursor Spectra}
In Fig.~\ref{fig:fp_VaryX} we show $f(x,p)$ calculated at various  positions, $x$, in the shock precursor as indicated. The top panel 
(Model~\eEE\ in Table~\ref{tab:param}) is 
for the Bohm limit, i.e., $\alfH=1$, while the bottom panel (Model~\fFF) has $\alfH=2$ with  $\pD =m_p c$.
The curve labeled $x=0$ in both panels displays the canonical $\sigma = 4.23$ spectral index above the quasi-thermal peak and below the cutoff caused by the upstream FEB. 
The large difference in transport length between Bohm and $\alfH=2$ shows up in the relation between $\Lfeb$ and $\pmax$. 
In the Bohm limit, $\LfebP=100$ yields $\pmax$ approximately 10 times as large as in the $\alfH=2$ example with $\LfebP=3\xx{6}$.

In both cases, the precursor spectra show that shocked particles 
scatter back upstream in an energy dependent fashion. It is noteworthy that this upstream transport  
readily occurs even though the speed difference between the accelerated particles and the bulk upstream flow is small, i.e., $u_0/c \simeq 0.995$ for $\gamZ=10$.
If particle scattering is assumed
as in equation~(\ref{eq:mfp}), the \rel\ nature of the flow has no qualitative effect on the particle transport, other than the effect on $\pFEB$ discussed in Section~\ref{sec:UM}.

Particles crossing the subshock at $x=0$ are highly anisotropic in the shock frame, but the anisotropy is essentially independent of momentum, as shown in Fig.~\ref{fig:PA_L1_L2} for four particle momentum ranges.
This  momentum independence of the \PAD, which is also independent of $\gamZ$ for fully \rel\ shocks, results in a power law with the canonical TP spectral index \citep[e.g.,][]{KW2005}.

Fig.~\ref{fig:PA_L1_L2} also shows that, to within statistics, the subshock crossing anisotropy is independent of the momentum dependence of $\Lmfp$ when the shock is unmodified.
We note that these \PADs\ are all normalized to the total injected number flux so they are absolutely normalized relative to each other.

\begin{figure}
\epsscale{1.0}
\plotone{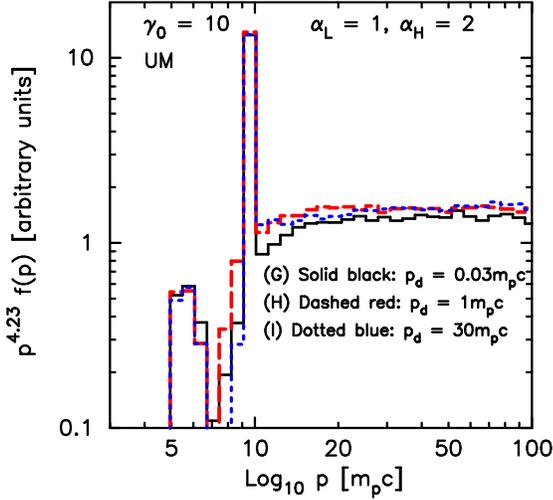}           
\caption{Low-momentum phase-space distributions for $\alfH=2$ with various $p_d$ as indicated. In all cases the \Facc\ is limited by setting $\pmax \gg 100\,m_pc$ and the spectra are calculated in the shock frame at $x=0$. While the differences in the spectra are mostly from statistical noise, the variation between the $\Pd=0.03\,m_p c$ spectrum and the other two just above the thermal peak may be real, albeit a consequence of the thermal leakage scheme assumed.
\label{fig:Inj}}
\end{figure}

\begin{figure}
\epsscale{0.95}
\plotone{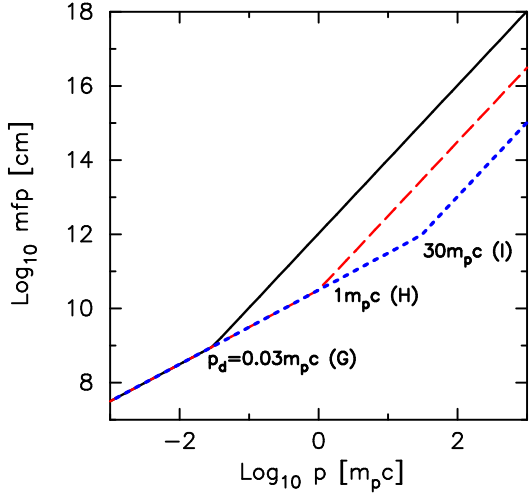}           
\caption{Mean free path for Models~\gGG, \hHH, and \iII\ as shown in 
Fig.~\ref{fig:Inj}.
\label{fig:Inj_mfp}}
\end{figure}

\subsection{Injection in UM Shocks} \label{sec:Inj}
In order to investigate how the  momentum dependence of $\Lmfp$ influences particle injection, we show in Fig.~\ref{fig:Inj} the low-momentum range of $f(p)$ for $\alfH=2$ for different values of $p_d$, as shown in 
Fig.~\ref{fig:Inj_mfp}. These examples all have $\alfL=\etamfp=1$, and we limit particle acceleration by setting an upstream FEB that is large enough in each case so $\pmax \gg 100\,m_pc$ for all $p_d$.\footnote{The value of $p_d$ modifies the scale of $\Lmfp$ and  thus will strongly influence $\pmax$. For a given shock size, smaller values of $p_d$ result in a lower $\pmax$.}
Concentrating on the low-energy part of $f(p)$, we see in 
Fig.~\ref{fig:Inj} that  $p_d$ has only a small effect on injection in these unmodified shocks.
For the factor of 1000 range in $p_d$, the normalization of $f(p)$ at $15-100\,m_pc$ varies by less than 15\%, most of which is statistical noise.
We show below that  $p_d$ has a much larger effect on \Facc\ in \SC, modified shocks.

It is interesting to note that there are peaks in Fig.~\ref{fig:Inj} below $p=\gamZ m_pc$ at $p \sim 6 m_pc$. These are produced by downstream particles making their first  crossing back into the upstream region. These particles  have momenta less than $p=\gamZ m_pc$ when measured in the shock frame. 
If the spectra in Fig.~\ref{fig:Inj} were plotted in the far upstream rest frame, peaks at $\sim \gamZ^2\,m_p c$ would be present.

\begin{figure}
\epsscale{0.95}
\plotone{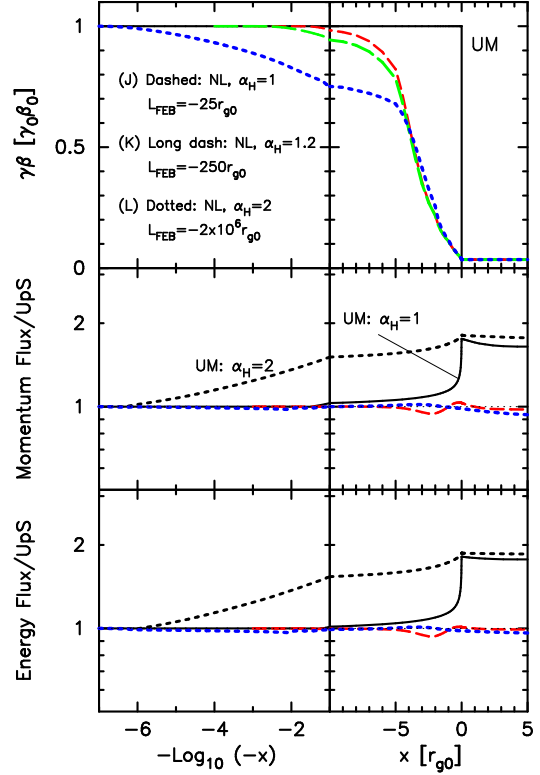}           
\caption{The top panels show the NL shock structure in terms of the inverse density, $\GBeta/(\GBetaZ)$, for shocks with different $\alfH$ and with upstream FEBs set to produce approximately the same $\pmax$, as shown in 
Fig.~\ref{fig:fp_same_pmax}.
The middle and bottom panels show the momentum and energy fluxes scaled to far upstream values.
The dashed (red, Model \jJJ), long-dashed (green, Model \kKK), 
and dotted (blue, Model \lLL) curves show \SC\ NL results, while the black curves show UM results, as indicated. The momentum and energy fluxes for the $\alfH=1.2$ case are not shown in the middle and bottom panels for clarity.
\label{fig:Prof_same_pmax}}
\end{figure}

\begin{figure}
\epsscale{0.95}
\plotone{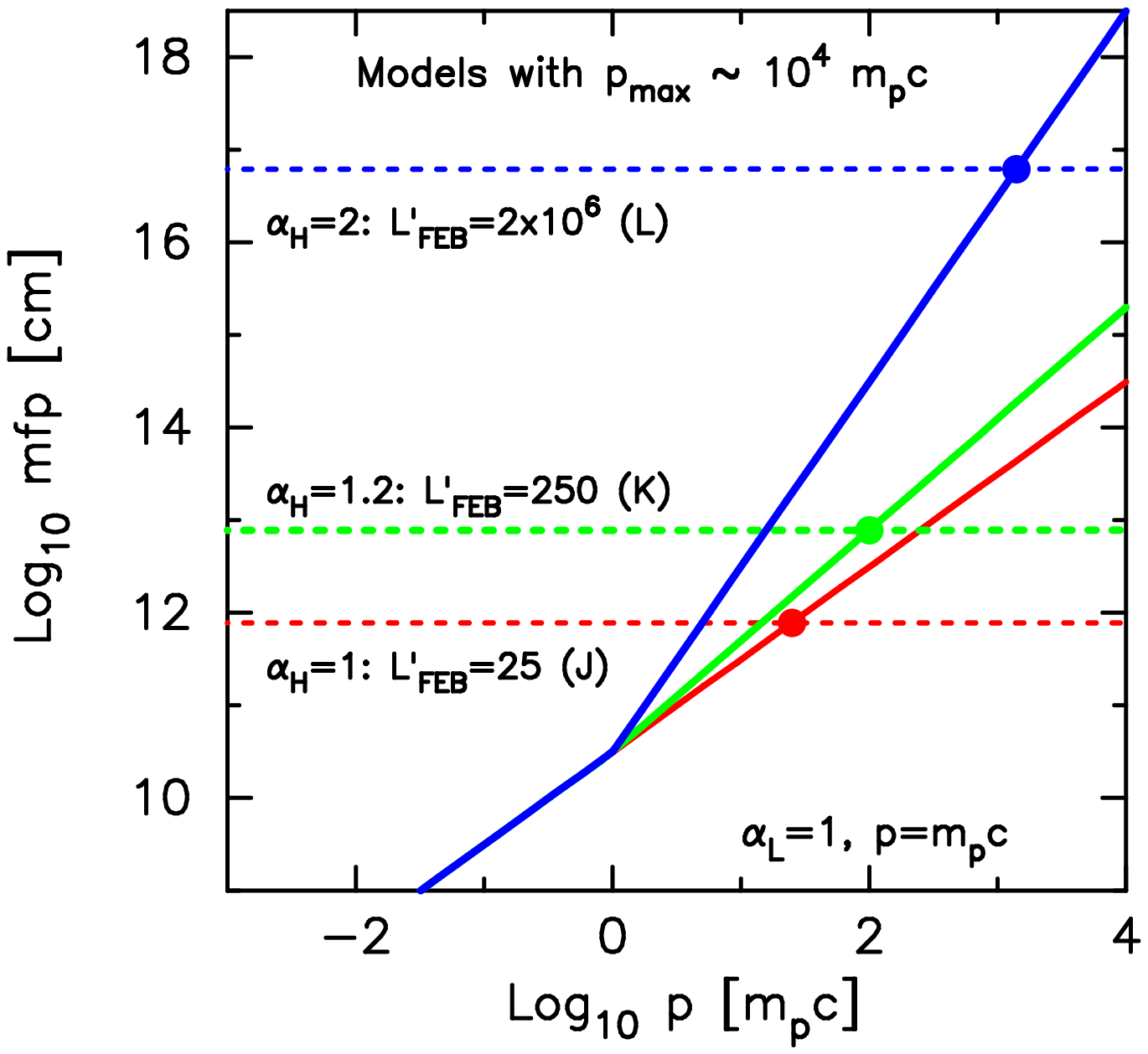}           
\caption{Mean free paths for the \SC\ shocks shown in 
Figs.~\ref{fig:Prof_same_pmax} and \ref{fig:fp_same_pmax}. 
The positions of the FEBs are shown and it is noted that the momenta at which $\Lmfp(\pFEB) \simeq |\Lfeb|$ (solid dots) are such that $\pFEB/\pmax$ is a strong increasing function of  $\alfH$.
\label{fig:same_mfp}}
\end{figure}

\begin{figure}
\epsscale{0.95}
\plotone{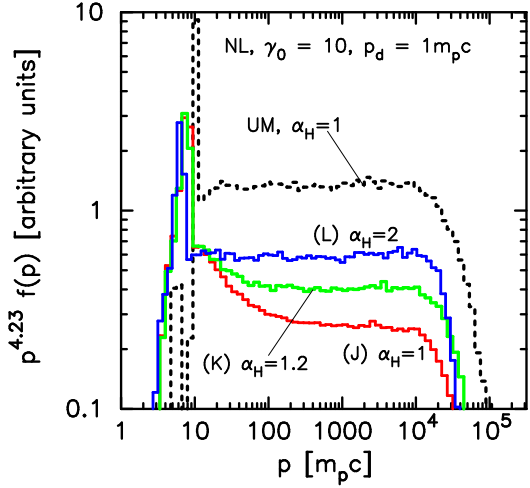}           
\caption{The lower three spectra are from the  \SC\ shocks shown in Fig.~\ref{fig:Prof_same_pmax} with different $\alfH$ as indicated. For comparison we show the spectrum  for an UM shock with $\alfH=1$ (dotted black curve). All spectra are calculated in the shock frame  at $x=0$ and multiplied by $p^{4.23}$. 
\label{fig:fp_same_pmax}}
\end{figure}

\subsection{Nonlinear Shocks}  \label{sec:NL}
We have shown that, other than a change in length scale and $\pmax$, the momentum dependence of $\Lmfp$ has no important influence on \Facc\ in UM shocks, at least within the assumptions and approximations of our \MC\ model.
We now investigate \NL\ shocks and show that the momentum dependence of $\Lmfp$ produces important differences beyond a simple scaling.

In Fig.~\ref{fig:Prof_same_pmax} we show \SC\ results for three values of $\alfH$ with $\alfL=p_d=1$ (models \jJJ, \kKK, and \lLL). In these examples, the upstream FEB has been set to produce approximately the same $\pmax \simeq 10^4\,m_p c$ for the three NL shocks (Fig~\ref{fig:fp_same_pmax}).
The top panels show 
inverse density profiles, $n_0/n(x) = \GBeta/(\GBetaZ)$, where $\gamma(x)$ and $\beta(x)$ are measured in the shock rest frame.
Here and elsewhere the subscript ``0" indicates far upstream values.
The mean free paths for these examples are shown in 
Fig.~\ref{fig:same_mfp}.

The \SC\ shock profiles shown as dashed ($\alfH=1$, red), 
long-dashed ($\alfH=1.2$, green), and dotted ($\alfH=2$, blue) curves in the top panels of Fig.~\ref{fig:Prof_same_pmax} conserve momentum and energy fluxes to within a few percent, as indicated in the middle and bottom panels. 
This is in contrast to the UM shocks (solid and dotted black curves) where the momentum and energy fluxes in the downstream region are well above the far upstream values.
Note the long extension of the momentum and energy fluxes into the precursor for the UM $\alfH=2$ case (black dotted curves), and the much shorter extension for the UM $\alfH=1$ case.

\begin{figure}
\epsscale{0.95}
\plotone{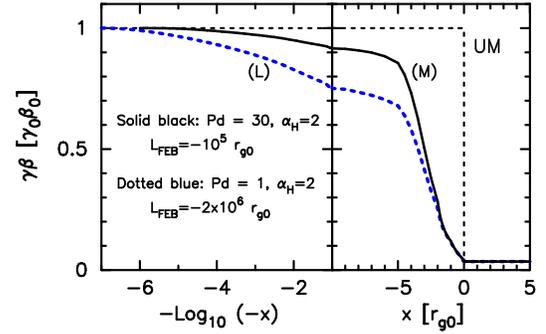}           
\caption{The NL shock structure in terms of the inverse density, $\GBeta/(\GBetaZ)$, for shocks with $\alfH=2$ and different $p_d$ as indicated. In both cases, the upstream FEB has been chosen to produce a $\pmax$ similar to the NL examples in Fig.~\ref{fig:fp_same_pmax}. 
While not shown, as in Fig.~\ref{fig:Prof_same_pmax}, the momentum and energy fluxes are conserved to within a few percent for the NL cases.
\label{fig:grip_pd30}}
\end{figure}

\begin{figure}
\epsscale{0.95}
\plotone{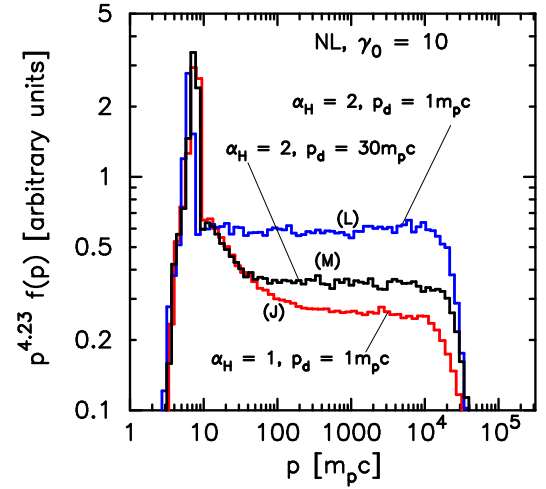}           
\caption{Phase-space distributions with various $\alfH$ and $p_d$, as indicated. The \jJJ\ and \lLL\ models are 
identical to those shown in Fig.~\ref{fig:fp_same_pmax}. Note the effect $p_d$ has on the \Facc\ efficiency in the top two curves both with $\alfH=2$. All spectra are calculated in the shock frame  at $x=0$.
\label{fig:fp_pd30}}
\end{figure}

In Fig.~\ref{fig:fp_same_pmax}, we show the shock-frame spectra calculated at $x=0$ for each model in Fig.~\ref{fig:Prof_same_pmax}.
The momentum dependence of $\Lmfp$ has a striking effect on $f(p)$. The Bohm limit case ($\alfH=1$, Model \jJJ) shows  the concave upward curvature characteristic of \NL\ \Facc\ in \nonrel\ shocks \citep[e.g.,][]{EE84} but this curvature is almost absent for 
$\alfH=2$ with $p_d=1$ (Model \lLL).
The NL spectra in 
Fig.~\ref{fig:fp_same_pmax} are absolutely normalized so they contain, within statistical limits, the same total particle and energy fluxes.
The lack of curvature for large $\alfH$ results in a significant increase in the acceleration efficiency and there is a slight shift of the thermal peak toward lower momentum to accommodate this change in efficiency. At $\sim  1000\,m_pc$ the NL $\alfH=2$ spectrum is $>$2 times as intense as the $\alfH=1$ 
spectrum.
 
The modification of the \Facc\ efficiency seen in 
Fig.~\ref{fig:fp_same_pmax} depends importantly on $p_d$ as well as $\alfH$. In Figs.~\ref{fig:grip_pd30} and \ref{fig:fp_pd30} we  compare results for $p_d=1$ (Model \lLL) and $30\,m_pc$ 
(Model \mMM) when $\alfH=2$. 
For the $p_d=30\,m_pc$ example (black solid curves in  
Figs.~\ref{fig:grip_pd30} and \ref{fig:fp_pd30}) we have adjusted the upstream FEB to give approximately the same $\pmax$ as the NL examples shown in Fig.~\ref{fig:fp_same_pmax}.
The spectrum with $p_d=30\,m_pc$ is now closer to the $p_d=m_pc$ case.

When upstream thermal particles cross the shock the first time and interact with the downstream plasma, they obtain a plasma frame momentum of $\simeq 10\,m_pc$.  For a $\Lmfp$ break at $p_d = 1\,m_pc$ between $\alfH=1$ and $\alfH = 2$, these upstream thermal particles obtain mean free paths roughly $500\,\rgz$ long.  This allows them to scatter far upstream from the subshock and treat much of the smooth shock transition as essentially unmodified.  
With $p_d=1\,m_pc$ this effect should increase smoothly as $\alfH$ goes from 1 to 2  since the scattering length is increasing but the shock structure isn't changing significantly  
(as seen in Fig.~\ref{fig:Prof_same_pmax}).
If the break point occurs above the downstream thermal peak at, for example $p_d=30\,m_pc$,  the \mfp\ for particles below this momentum will be much less. 
In Fig.~\ref{fig:fp_pd30}, the $\alfH=1$ ($p_d=m_pc$) and $\alfH=2$ ($p_d=30\,m_pc$) spectra behave similarly up to the separation momentum  at $p = 30\,m_pc$. Particles with $p < p_d$ feel the smooth shock structure and obtain the concave shape. For $p>p_d$, the $\alfH=1$ and $\alfH=2$ distributions separate.
As Fig.~\ref{fig:fp_pd30} clearly shows, the form of $\alfP$ will influence NL \Facc\ significantly.

In Table~\ref{tab:param}, we define the particle acceleration efficiency, $\EffDSA$, for our NL $\gamZ=10$  models as the fraction of energy flux above 1\,TeV, measured in the shock frame. Since we are not attempting to fit a particular object with realistically derived parameters, $\EffDSA$ is intended only as a measure of the relative shock acceleration efficiency for the various models. Table~\ref{tab:param} shows that Model \lLL\ with $\alfH=2$ puts approximately twice the energy into $>$TeV protons as does Model \jJJ\ with $\alfH=1$.

\begin{figure}
\epsscale{0.95}
\plotone{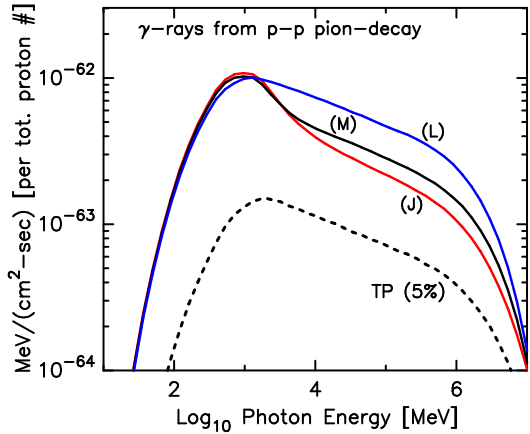}           
\caption{Local plasma frame \gamray\ emission due to $\pi^0$-decay produced in inealstic collisions of accelerated protons for the models  shown in  Fig.~\ref{fig:fp_pd30}. 
Model \jJJ\ has $\alfH=1$, model \lLL\ has $\alfH=2$ with $\Pd=m_p c$, and model \mMM\ has $\alfH=2$ with $\Pd=30\,m_p c$.
These spectra have an arbitrary absolute scale but are normalized to total proton number in each emitting region to isolate the effects from a varying $p$-dependence for $\Lmfp$. The \pion\ emission is calculated using parameterizations 
given by \citet{Kamae06,Kamae2007} and \citet{Kelner2009}.
The dotted curve is the TP \gamray\ emission for model \jJJ\ parameters assuming that 5\% of the proton energy flux goes into Fermi accelerated CRs.
\label{fig:pion}}
\end{figure}

\subsection{Pion-decay Emission}
In Fig.~\ref{fig:pion} we show the $\pi^0$-decay \gamray\ emission from the protons shown in Fig.~\ref{fig:fp_pd30}. The density of the ambient material for the proton-proton interactions is determined \SCly\ within the NL shock structure, i.e., from $\GBeta$ in 
Figs.~\ref{fig:Prof_same_pmax} and \ref{fig:grip_pd30}.
The spectra in Fig.~\ref{fig:pion} are normalized to the number of protons in the shock acceleration region and thus emphasize the NL  effects from shock smoothing. 
As described in detail  in \citet{WarrenEllison2015}, the photon emission depends on the number of particles in the region between the upstream and downstream FEBs. For a given $\pmax$, the spatial extent of the shock, and the absolute normalization of the radiation, depend strongly on both $\alfH$ and $\Pd$. 

Fig.~\ref{fig:pion} shows that in addition to this simple scaling, NL effects from shock smoothing can produce a significant enhancement of the photon emission from that expected from the Bohm limit.
Fig.~\ref{fig:pion} also shows that the spectral shape of the \pion\ emission in the 1-100\,GeV range is modified substantially by NL effects.
The dotted curve in the figure is the TP result where we have arbitrarily assumed that 5\% of the proton energy flux is put into Fermi accelerated particles. At 100\,GeV, the photon emission from the TP result  is a factor of three or more below any of the NL models.
Note that for a given $\pmax$, the TP result does not depend on $\alfH$ or $\Pd$.

For simplicity, we have not included the acceleration of electrons or the leptonic emission here but similar effects can be 
expected \citep[see][for a discussion of electron 
acceleration and emission]{WarrenEllison2015}.
We believe the NL effects from the momentum dependence of $\Lmfp$ may be important for a detailed modeling of objects containing \rel\ shocks, such as GRBs, but this is beyond the scope of this paper and is left for future work.

\begin{figure}
\epsscale{0.95}
\plotone{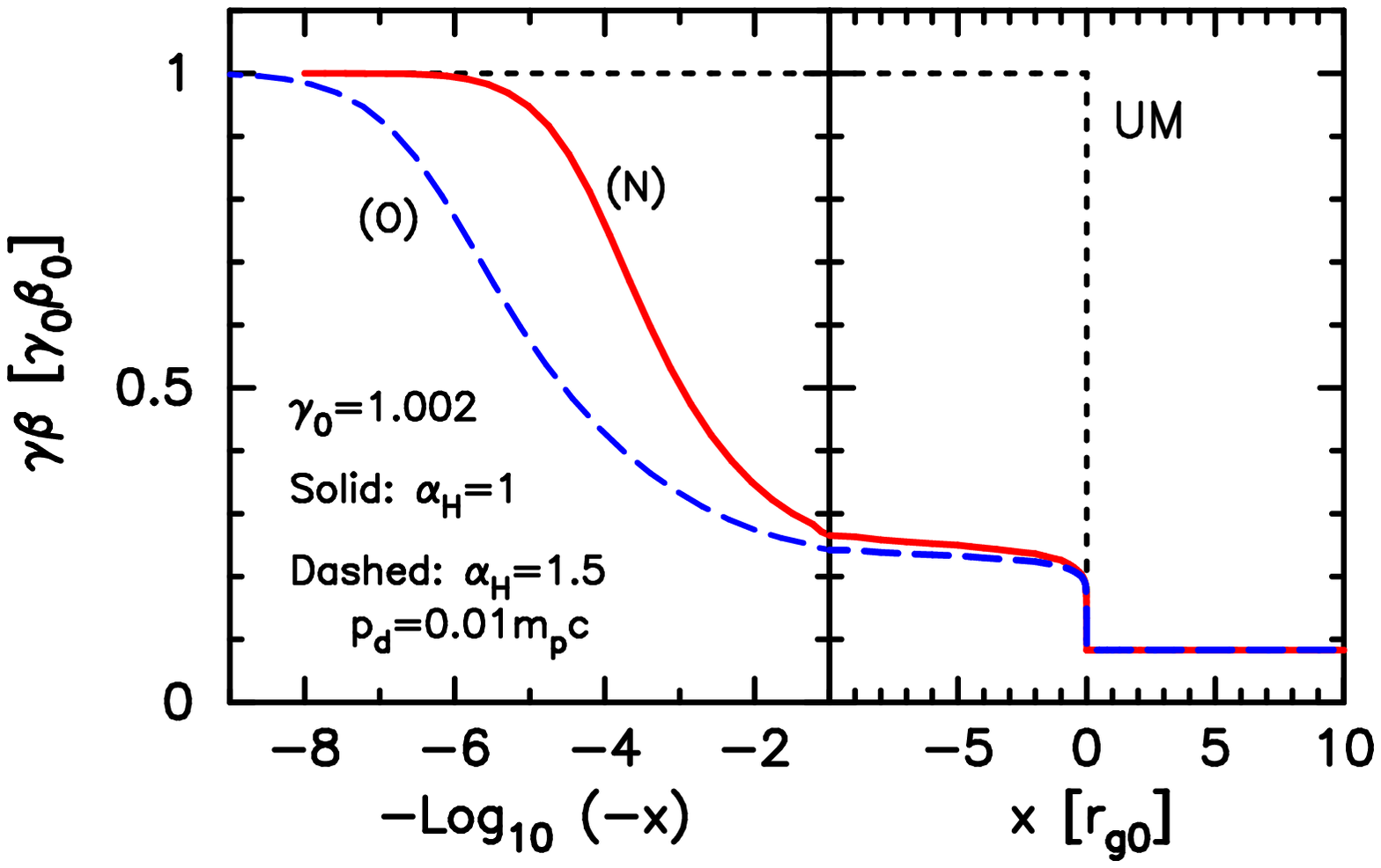}           
\caption{Shock structure for \nonrel\ shocks (i.e., $\gamZ=1.002$) with two values of $\alfH$, as indicated. 
For the $\alfH=1.5$ example, $p_d=0.01\,m_pc$. The upstream FEB has been set to produce approximately the same $\pmax$, as shown in 
Fig.~\ref{fig:fp_nonrel}.
As is well known for \nonrel\ shocks undergoing efficient shock acceleration, the overall shock compression ratio must increase above the \RH\ value \citep[e.g.,][]{JE91,BE99}. For these examples, $\Rtot \simeq 12$. 
\label{fig:grid_nonrel}}
\end{figure}

\begin{figure}
\epsscale{0.95}
\plotone{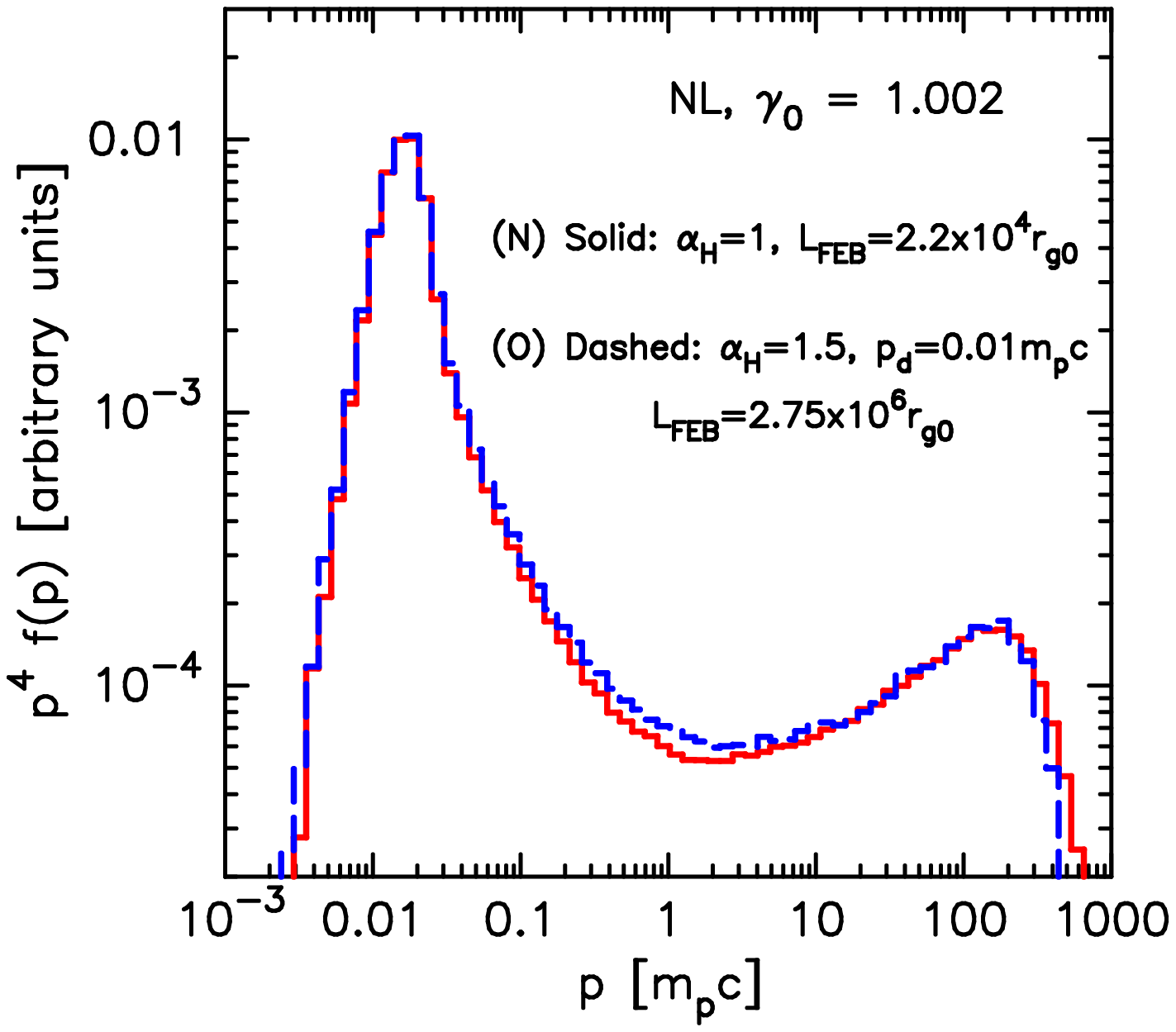}           
\caption{Shock frame distributions calculated at $x=0$ for the Model \nNN\ and \oOO\ shocks shown in Fig.~\ref{fig:grid_nonrel}. The \SC\ shock structure adjusts to produce approximately identical CR distributions despite the large change in precursor scale produced by the change in the $p$-dependence of $\Lmfp$. These two examples are extremely efficient with $\EffDSA \sim 0.8$, where $\EffDSA$ is defined as the fraction of proton energy flux above  $m_pc$ for \nonrel\ shocks, as in Table~\ref{tab:param}.
\label{fig:fp_nonrel}}
\end{figure}

\subsection{Nonlinear, \Nonrel\ Shocks} \label{sec:NR}
To confirm  that in parallel, \nonrel\ shocks, our \MC\ code gives results that are independent of the momentum dependence of $\Lmfp$, we show NL shock structures and spectra for two values of $\alfH$ 
in 
Figs.~\ref{fig:grid_nonrel} and \ref{fig:fp_nonrel} for shocks with $\gamZ=1.002$ ($u_0 \simeq 1.9\xx{4}$\,\kmps). The $\alfH=1.5$ example has $p_d=0.01\,m_pc$.
As in Fig.~\ref{fig:Prof_same_pmax}, the upstream FEB has been adjusted to give approximately the same $\pmax$ for each case.
Fig.~\ref{fig:grid_nonrel} shows the large difference in precursor scale caused by $\alfH$ but Fig.~\ref{fig:fp_nonrel} shows that the CR distribution functions are almost identical. 
In contrast to \rel\ shocks, 
the momentum dependence of $\Lmfp$ has no effect on \NL\ \Facc\ in \nonrel\ shocks, other than the change in scale,  as long as they are parallel and  simple thermal leakage injection is assumed.

The two examples in these figures also confirm that, to within statistics, $\pFEB = \pmax$, where $\pFEB$ is defined in 
equation~(\ref{eq:Pfeb}). 
For more details on NL \Facc\  in \nonrel\ shocks, the reader is directed to \citet{BE99,MD2001,CKVJ2010,Bykov3inst2014} and references therein.

\newlistDE

\section{Conclusions}
Using a generalized form for the \PAS\ \mfp, $\Lmfp(p) \propto p^{\alfP}$, in a 
kinematic \mc\ model making specific  scattering assumptions, we  have investigated how the momentum dependence of $\Lmfp$ influences \Facc\ in \rel\ shocks.
Our main results are summarized below.

\listDE 
For a given shock size\footnote{The examples presented here assume 
steady-state conditions so only the size of the shock system is considered. In a time-dependent model, the acceleration time to a particular momentum would also scale with the momentum dependence of $\Lmfp$.} 
the increase in scattering 
length produced by a strong momentum dependence for $\Lmfp$ dramatically reduces the maximum energy a given shock can produce (Fig.~\ref{fig:fp_UM}).

\listDE
Superthermal particles can readily move into the shock precursor 
(Fig.~\ref{fig:fp_VaryX}) as for \nonrel\ flows, but the 
particle transport no longer obeys the simple $D/u_0$ relation for the UM precursor diffusion length as in \nonrel\ shocks 
(i.e., equation~\ref{eq:Lfeb}): particles must obtain a greater momentum to reach an upstream FEB than $D/u_0$ suggests (i.e., equation~\ref{eq:Pfeb}). 
Precursor transport approaches the \nonrel\ expression as the strength of the momentum dependence of $\Lmfp$ increases.

\listDE 
In unmodified shocks (those that ignore the backreaction of accelerated particles), adjusting $\Lmfp(p)$ has no important effects beyond the change in $\pmax$.  Both the power-law spectral index 
(Fig.~\ref{fig:fp_UM}) and the thermal leakage injection efficiency (Fig.~\ref{fig:Inj}) are  nearly independent of $\Lmfp$ as long as \TP\ conditions are assumed. 

\listDE
Once the nonlinear interaction between the bulk flow and the accelerated particles is considered, both the injection efficiency and accelerated particle spectral shape depend  significantly on the momentum dependence assumed for $\Lmfp(p)$ 
(Figs.~\ref{fig:Prof_same_pmax} and \ref{fig:fp_same_pmax}).

\listDE
If the upstream FEB is adjusted to produce a similar maximum CR energy,  NL shocks with a momentum dependence for $\Lmfp(p)$ stronger than Bohm (i.e., $\Lmfp \propto p^{\alfP}$ with $\alfP > 1$), inject and accelerate CRs more efficiently than in the Bohm limit. This effect comes about solely from the transport properties as determined by equation~(\ref{eq:mfp})
and does not occur in NL, \nonrel\ shocks or in \rel\ shocks where the backreaction of CRs on the shock structure is ignored.

\listDE
The increase in \Facc\ efficiency produces a corresponding increase in the \gamray\ emissivity for $\alfP>1$ for \SC\ models  (Fig.~\ref{fig:pion}). 
For the parameters used here, NL effects can produce a  factor of three enhancement in the \pion\  flux between $\alfP=1$ and $2$ 
in the $10-100$\,GeV range.
In this energy range, the \pion\ spectrum can be noticeably  harder for large $\alfP$ compared to the Bohm limit.
An even larger difference in \gamray\ emissivity is predicted between TP approximations and NL shocks with thermal leakage injection.

\PIC\  simulations have shown that unmagnetized \rel\ shocks can be efficient particle accelerators \citep[e.g.,][]{SSA2013} regardless of magnetic field geometry. 
The unmagnetized condition should apply for GRB external shocks producing afterglow emission, the early-time blast waves for particularly powerful supernovae, and possibly other sources containing \rel\ shocks.
If \Facc\ is efficient in these cases \citep[and the $\sim 10\%$ efficiencies  reported by][lower limits considering the limited box size and run times of the PIC simulations, are in line with what we obtain in our NL models]{SSA2013}  the NL effects of this acceleration must be considered for a \SC\ description of the shock formation and structure, magnetic turbulence generation, and CR production.

The self-generation of magnetic turbulence, and the particle scattering that results from it,  are critical and poorly understood components of collisionless shock formation and \Facc.
A great deal of work has been done studying the generation of magnetic  turbulence in \TP, \rel\ shocks with \SA\ and hydrodynamic techniques 
\citep[e.g.,][]{PLM2009,DelaneyEtal2012,GuidorziEtal2014,Bosch2015}, but the \NL\ problem with efficient \Facc\ has not yet been adequately addressed  with these techniques.

Indeed, at present the full \SC\ problem can only be solved with PIC simulations, and an important result of that work is the demonstration that the Weibel instability produces short-wavelength turbulence close to the subshock layer, i.e., within a few hundred 
ion skin depths \citep[see Eq. 6 in][]{SSA2013}.
Particles moving through turbulence with scales much shorter than their gyroradius will scatter with $\Lmfp \propto p^2$, very different from the generally assumed Bohm diffusion.
\PIC\ simulations, however,  are computationally intensive and are limited in box size, run time, and dimensionality 
\citep[see][]{JJB98,VBE2008}. It is still an open question whether instabilities can be self-generated further into the shock precursor on scales comparable to the particle gyroradius 
\citep[e.g.,][]{SagiNakar2012,LPGP2014}.

If \rel\ shocks in astrophysics do, in fact, efficiently Fermi accelerate CRs 
to ultra-high energies \citep[e.g.,][]{Globus2015}, they must somehow produce long-wavelength magnetic turbulence in the shock precursor and computationally fast techniques are required to model the resultant \Facc. 
The \mc\ model we have presented here concentrates on the NL kinematics and parameterizes the particle transport, assuming it can be produced well into the precursor.
The transport model is more general than the 
diffusion-advection equation widely used in \Facc\ and can account for an arbitrary particle anisotropy.

Of course, many important aspects of \rel\ shocks remain in question and we make no claim that our \mc\ simulation is a first-principles calculation. The models presented here, for instance, assume a plane shock and have no position dependence for $\Lmfp(p)$. While PIC simulations show that magnetic turbulence is produced near the subshock and in the shock precursor, it may decay rapidly downstream from the subshock in a wavelength-dependent fashion. Using fundamental PIC results to guide simple parameterizations, future work will include a position and 
momentum-dependent \mfp, as well as the calculation of plasma instabilities \SCly\ within the \mc\ code, as has been done for \nonrel\ shocks 
\citep[e.g.,][]{VEB2006,VBE2009,BOE2011,Bykov3inst2014}.
Anisotropic scattering can also be modeled using PIC results to guide the parameterization.

A further extension of this work is application to the evolving conditions of a GRB afterglow.  Significant progress has already been made in this area, starting shortly after the first afterglow observations with the analytic work of \citet{SPN1998} and references therein.  In recent years the traditional power-law electron distribution has been coupled to relativistically correct hydrodynamics calculations 
\citep[e.g.,][]{Leventis2012,vanEertenMacFadyen2013}, allowing for rapid estimation of key GRB parameters from the afterglow light curves and spectra.  However, a key feature of these models is their assumption that non-thermal electrons form a single power law with constant spectral index.  We have shown here and in earlier work \citep{EWB2013,WarrenEllison2015} that the acceleration process is subject to many uncertainties, and that a simple power law may be insufficient to describe the particle distribution at any particular instant in time, let alone for a substantial fraction of the afterglow.  A preliminary treatment of nonlinear DSA in the context of afterglows was presented in \citet{Warren2015dis}, and further study is planned.

\acknowledgments The authors acknowledge useful discussions with 
Martin Lemoine and Guy Pelletier. 
D.C.E. and D.C.W. acknowledge support from NASA
grant NNX11AE03G. 
Part of this work was performed at the Aspen Center for Physics, which is supported by National Science Foundation grant PHY-1066293.

\bibliographystyle{aa} 
\bibliography{bib_DCE}

\begin{thebibliography}{58}
\expandafter\ifx\csname natexlab\endcsname\relax\def\natexlab#1{#1}\fi

\bibitem[{{Achterberg} {et~al.}(2001){Achterberg}, {Gallant}, {Kirk}, \&
  {Guthmann}}]{Achterberg2001}
{Achterberg}, A., {Gallant}, Y.~A., {Kirk}, J.~G., \& {Guthmann}, A.~W. 2001,
  \mnras, 328, 393

\bibitem[{{Bednarz} \& {Ostrowski}(1998)}]{BO98}
{Bednarz}, J. \& {Ostrowski}, M. 1998, Physical Review Letters, 80, 3911

\bibitem[{Berezhko \& Ellison(1999)}]{BE99}
Berezhko, E.~G. \& Ellison, D.~C. 1999, ApJ, 526, 385

\bibitem[{{Bosch-Ramon}(2015)}]{Bosch2015}
{Bosch-Ramon}, V. 2015, \aap, 575, A109

\bibitem[{{Bykov} {et~al.}(2012){Bykov}, {Gehrels}, {Krawczynski}, {Lemoine},
  {Pelletier}, \& {Pohl}}]{BykovEtal2012}
{Bykov}, A., {Gehrels}, N., {Krawczynski}, H., {et~al.} 2012, \ssr, 173, 309

\bibitem[{{Bykov} {et~al.}(2014){Bykov}, {Ellison}, {Osipov}, \&
  {Vladimirov}}]{Bykov3inst2014}
{Bykov}, A.~M., {Ellison}, D.~C., {Osipov}, S.~M., \& {Vladimirov}, A.~E. 2014,
  ApJ, 789, 137

\bibitem[{{Bykov} {et~al.}(2011){Bykov}, {Osipov}, \& {Ellison}}]{BOE2011}
{Bykov}, A.~M., {Osipov}, S.~M., \& {Ellison}, D.~C. 2011, \mnras, 410, 39

\bibitem[{{Caprioli} {et~al.}(2010){Caprioli}, {Kang}, {Vladimirov}, \&
  {Jones}}]{CKVJ2010}
{Caprioli}, D., {Kang}, H., {Vladimirov}, A.~E., \& {Jones}, T.~W. 2010,
  \mnras, 407, 1773

\bibitem[{{Casse} {et~al.}(2013){Casse}, {Marcowith}, \& {Keppens}}]{Casse2013}
{Casse}, F., {Marcowith}, A., \& {Keppens}, R. 2013, \mnras, 433, 940

\bibitem[{{Delaney} {et~al.}(2012){Delaney}, {Dempsey}, {Duffy}, \&
  {Downes}}]{DelaneyEtal2012}
{Delaney}, S., {Dempsey}, P., {Duffy}, P., \& {Downes}, T.~P. 2012, \mnras,
  420, 3360

\bibitem[{{Double} {et~al.}(2004){Double}, {Baring}, {Jones}, \&
  {Ellison}}]{DoubleEtal2004}
{Double}, G.~P., {Baring}, M.~G., {Jones}, F.~C., \& {Ellison}, D.~C. 2004,
  \apj, 600, 485

\bibitem[{{Drury}(2011)}]{Drury2011}
{Drury}, L.~O. 2011, \mnras, 415, 1807

\bibitem[{{Ellison} \& {Double}(2002)}]{ED2002}
{Ellison}, D.~C. \& {Double}, G.~P. 2002, Astroparticle Physics, 18, 213

\bibitem[{{Ellison} \& {Double}(2004)}]{ED2004}
{Ellison}, D.~C. \& {Double}, G.~P. 2004, Astroparticle Physics, 22, 323

\bibitem[{{Ellison} \& {Eichler}(1984)}]{EE84}
{Ellison}, D.~C. \& {Eichler}, D. 1984, \apj, 286, 691

\bibitem[{{Ellison} {et~al.}(1990){Ellison}, {Jones}, \& {Reynolds}}]{EJR90}
{Ellison}, D.~C., {Jones}, F.~C., \& {Reynolds}, S.~P. 1990, \apj, 360, 702

\bibitem[{{Ellison} {et~al.}(2007){Ellison}, {Patnaude}, {Slane}, {Blasi}, \&
  {Gabici}}]{EPSBG2007}
{Ellison}, D.~C., {Patnaude}, D.~J., {Slane}, P., {Blasi}, P., \& {Gabici}, S.
  2007, \apj, 661, 879

\bibitem[{{Ellison} {et~al.}(2013){Ellison}, {Warren}, \& {Bykov}}]{EWB2013}
{Ellison}, D.~C., {Warren}, D.~C., \& {Bykov}, A.~M. 2013, ApJ, 776, 46

\bibitem[{{Globus} {et~al.}(2015){Globus}, {Allard}, {Mochkovitch}, \&
  {Parizot}}]{Globus2015}
{Globus}, N., {Allard}, D., {Mochkovitch}, R., \& {Parizot}, E. 2015, \mnras,
  451, 751

\bibitem[{{Guidorzi} {et~al.}(2014){Guidorzi}, {Mundell}, {Harrison},
  {Margutti}, {Sudilovsky}, {Zauderer}, {Kobayashi}, {Cucchiara}, {Melandri},
  {Pandey}, {Berger}, {Bersier}, {D'Elia}, {Gomboc}, {Greiner}, {Japelj},
  {Kopa{\v c}}, {Kumar}, {Malesani}, {Mottram}, {O'Brien}, {Rau}, {Smith},
  {Steele}, {Tanvir}, \& {Virgili}}]{GuidorziEtal2014}
{Guidorzi}, C., {Mundell}, C.~G., {Harrison}, R., {et~al.} 2014, \mnras, 438,
  752

\bibitem[{{Haugb{\o}lle}(2011)}]{Haugbolle2011}
{Haugb{\o}lle}, T. 2011, \apjl, 739, L42

\bibitem[{{Jokipii}(1972)}]{Jokipii72}
{Jokipii}, J.~R. 1972, \apj, 172, 319

\bibitem[{{Jones} \& {Ellison}(1991)}]{JE91}
{Jones}, F.~C. \& {Ellison}, D.~C. 1991, Space Science Reviews, 58, 259

\bibitem[{{Jones} {et~al.}(1998){Jones}, {Jokipii}, \& {Baring}}]{JJB98}
{Jones}, F.~C., {Jokipii}, J.~R., \& {Baring}, M.~G. 1998, \apj, 509, 238

\bibitem[{{Kamae} {et~al.}(2006){Kamae}, {Karlsson}, {Mizuno}, {Abe}, \&
  {Koi}}]{Kamae06}
{Kamae}, T., {Karlsson}, N., {Mizuno}, T., {Abe}, T., \& {Koi}, T. 2006, \apj,
  647, 692

\bibitem[{{Kamae} {et~al.}(2007){Kamae}, {Karlsson}, {Mizuno}, {Abe}, \&
  {Koi}}]{Kamae2007}
{Kamae}, T., {Karlsson}, N., {Mizuno}, T., {Abe}, T., \& {Koi}, T. 2007, \apj,
  662, 779

\bibitem[{{Kato} \& {Takabe}(2008)}]{KT2008}
{Kato}, T.~N. \& {Takabe}, H. 2008, \apjl, 681, L93

\bibitem[{{Kelner} {et~al.}(2009){Kelner}, {Aharonian}, \&
  {Bugayov}}]{Kelner2009}
{Kelner}, S.~R., {Aharonian}, F.~A., \& {Bugayov}, V.~V. 2009, \prd, 79, 039901

\bibitem[{{Keshet} {et~al.}(2009){Keshet}, {Katz}, {Spitkovsky}, \&
  {Waxman}}]{KeshetEtal2009}
{Keshet}, U., {Katz}, B., {Spitkovsky}, A., \& {Waxman}, E. 2009, \apjl, 693,
  L127

\bibitem[{{Keshet} \& {Waxman}(2005)}]{KW2005}
{Keshet}, U. \& {Waxman}, E. 2005, Physical Review Letters, 94, 111102

\bibitem[{{Kulkarni} {et~al.}(1999){Kulkarni}, {Djorgovski}, {Odewahn},
  {Bloom}, {Gal}, {Koresko}, {Harrison}, {Lubin}, {Armus}, {Sari},
  {Illingworth}, {Kelson}, {Magee}, {van Dokkum}, {Frail}, {Mulchaey},
  {Malkan}, {McClean}, {Teplitz}, {Koerner}, {Kirkpatrick}, {Kobayashi},
  {Yadigaroglu}, {Halpern}, {Piran}, {Goodrich}, {Chaffee}, {Feroci}, \&
  {Costa}}]{Kulkarni1999}
{Kulkarni}, S.~R., {Djorgovski}, S.~G., {Odewahn}, S.~C., {et~al.} 1999, \nat,
  398, 389

\bibitem[{{Lemoine} \& {Pelletier}(2010)}]{LP2010}
{Lemoine}, M. \& {Pelletier}, G. 2010, \mnras, 402, 321

\bibitem[{{Lemoine} {et~al.}(2014){Lemoine}, {Pelletier}, {Gremillet}, \&
  {Plotnikov}}]{LPGP2014}
{Lemoine}, M., {Pelletier}, G., {Gremillet}, L., \& {Plotnikov}, I. 2014,
  \mnras, 440, 1365

\bibitem[{{Leventis} {et~al.}(2012){Leventis}, {van Eerten}, {Meliani}, \&
  {Wijers}}]{Leventis2012}
{Leventis}, K., {van Eerten}, H.~J., {Meliani}, Z., \& {Wijers}, R.~A.~M.~J.
  2012, \mnras, 427, 1329

\bibitem[{{Malkov} \& {Drury}(2001)}]{MD2001}
{Malkov}, M.~A. \& {Drury}, L. 2001, Reports of Progress in Physics, 64, 429

\bibitem[{{Marcowith} {et~al.}(2006){Marcowith}, {Lemoine}, \&
  {Pelletier}}]{marcowith06}
{Marcowith}, A., {Lemoine}, M., \& {Pelletier}, G. 2006, \aap, 453, 193

\bibitem[{{Morlino} {et~al.}(2009){Morlino}, {Amato}, \& {Blasi}}]{MAB2009}
{Morlino}, G., {Amato}, E., \& {Blasi}, P. 2009, \mnras, 392, 240

\bibitem[{{Nishikawa} {et~al.}(2009){Nishikawa}, {Niemiec}, {Hardee},
  {Medvedev}, {Sol}, {Mizuno}, {Zhang}, {Pohl}, {Oka}, \&
  {Hartmann}}]{Nishikawa2009}
{Nishikawa}, K.-I., {Niemiec}, J., {Hardee}, P.~E., {et~al.} 2009, \apjl, 698,
  L10

\bibitem[{{Ostrowski}(1991)}]{Ostrowski1991}
{Ostrowski}, M. 1991, \mnras, 249, 551

\bibitem[{{Pelletier} {et~al.}(2009){Pelletier}, {Lemoine}, \&
  {Marcowith}}]{PLM2009}
{Pelletier}, G., {Lemoine}, M., \& {Marcowith}, A. 2009, \mnras, 393, 587

\bibitem[{{Plotnikov} {et~al.}(2011){Plotnikov}, {Pelletier}, \&
  {Lemoine}}]{PPL2011}
{Plotnikov}, I., {Pelletier}, G., \& {Lemoine}, M. 2011, \aap, 532, A68

\bibitem[{{Plotnikov} {et~al.}(2013){Plotnikov}, {Pelletier}, \&
  {Lemoine}}]{PPL2013}
{Plotnikov}, I., {Pelletier}, G., \& {Lemoine}, M. 2013, \mnras, 430, 1280

\bibitem[{{Rabinak} {et~al.}(2011){Rabinak}, {Katz}, \& {Waxman}}]{RKW2011}
{Rabinak}, I., {Katz}, B., \& {Waxman}, E. 2011, \apj, 736, 157

\bibitem[{{Reville} \& {Bell}(2014)}]{RevilleBell2014}
{Reville}, B. \& {Bell}, A.~R. 2014, \mnras, 439, 2050

\bibitem[{{Sagi} \& {Nakar}(2012)}]{SagiNakar2012}
{Sagi}, E. \& {Nakar}, E. 2012, \apj, 749, 80

\bibitem[{{Sari} {et~al.}(1998){Sari}, {Piran}, \& {Narayan}}]{SPN1998}
{Sari}, R., {Piran}, T., \& {Narayan}, R. 1998, \apjl, 497, L17

\bibitem[{{Schlickeiser}(2015)}]{Schlickeiser2015}
{Schlickeiser}, R. 2015, \apj, 809, 124

\bibitem[{{Sironi} \& {Spitkovsky}(2011)}]{SironiSpit2011}
{Sironi}, L. \& {Spitkovsky}, A. 2011, \apj, 726, 75

\bibitem[{{Sironi} {et~al.}(2013){Sironi}, {Spitkovsky}, \& {Arons}}]{SSA2013}
{Sironi}, L., {Spitkovsky}, A., \& {Arons}, J. 2013, ApJ, 771, 54

\bibitem[{{Spitkovsky}(2008)}]{Spitkovsky2008}
{Spitkovsky}, A. 2008, \apjl, 673, L39

\bibitem[{{Summerlin} \& {Baring}(2012)}]{SummerlinBaring2012}
{Summerlin}, E.~J. \& {Baring}, M.~G. 2012, \apj, 745, 63

\bibitem[{{Trattner} {et~al.}(2013){Trattner}, {Allegrini}, {Dayeh}, {Funsten},
  {Fuselier}, {Heirtzler}, {Janzen}, {Kucharek}, {McComas}, {M{\"o}bius},
  {Moore}, {Petrinec}, {Reisenfeld}, {Schwadron}, \& {Wurz}}]{Trattner2013}
{Trattner}, K.~J., {Allegrini}, F., {Dayeh}, M.~A., {et~al.} 2013, Journal of
  Geophysical Research (Space Physics), 118, 4425

\bibitem[{{van Eerten} \& {MacFadyen}(2013)}]{vanEertenMacFadyen2013}
{van Eerten}, H. \& {MacFadyen}, A. 2013, \apj, 767, 141

\bibitem[{{Vladimirov} {et~al.}(2006){Vladimirov}, {Ellison}, \&
  {Bykov}}]{VEB2006}
{Vladimirov}, A., {Ellison}, D.~C., \& {Bykov}, A. 2006, ApJ, 652, 1246

\bibitem[{{Vladimirov} {et~al.}(2008){Vladimirov}, {Bykov}, \&
  {Ellison}}]{VBE2008}
{Vladimirov}, A.~E., {Bykov}, A.~M., \& {Ellison}, D.~C. 2008, \apj, 688, 1084

\bibitem[{{Vladimirov} {et~al.}(2009){Vladimirov}, {Bykov}, \&
  {Ellison}}]{VBE2009}
{Vladimirov}, A.~E., {Bykov}, A.~M., \& {Ellison}, D.~C. 2009, \apjl, 703, L29

\bibitem[{{Warren}(2015)}]{Warren2015dis}
{Warren}, D.~C. 2015, PhD thesis, North Carolina State University

\bibitem[{{Warren} {et~al.}(2015){Warren}, {Ellison}, {Bykov}, \&
  {Lee}}]{WarrenEllison2015}
{Warren}, D.~C., {Ellison}, D.~C., {Bykov}, A.~M., \& {Lee}, S.-H. 2015,
  \mnras, 452, 431

\end{thebibliography}

\clearpage


\begin{table}
\begin{center}
\caption{Model Parameters.}
\label{tab:param}
\renewcommand{\arraystretch}{1.5} 
\begin{tabular}{crrrrrrrrrrrrrr}
\tableline \tableline
\\
%
Model\tablenotemark{a}\tablenotetext{1}{All models have $\alfL=1$, $n_p=1$\,\pcc, far upstream proton temperature $T_p=10^6$\,K, $B_0=10^{-4}$\,G, and $\etamfp=1$. While electrons are not modeled explicitly, they are included in all models with $n_e=n_p$ and $T_e=T_p$ in the determination of the shock compression ratio.}
&Type\tablenotemark{b}\tablenotetext{2}{In the \NL\ (NL) models the shock structure is determined \SCly. The \UM\ (UM) models have a discontinuous shock structure with no shock smoothing.}
&$\gamZ$ 
&$\alfH$ 
&$\pD$\,[$m_pc$]
&$|\LfebUpS|$\,[$\rgz$]
&$\EffDSA$\tablenotemark{c}\tablenotetext{3}{This is a measure of the shock acceleration efficiency for NL models. For the $\gamZ=10$ examples, $\EffDSA$ is the fraction of total energy flux placed in protons with energies 
above 1\,TeV, as measured in the shock frame at $x=0$. For the $\gamZ=1.002$ models, it is the energy flux fraction in protons with $p>m_p c$.} 
\\
\tableline
\aaT 
&UM
&10 
&1
&1 
&$10^3$
&\dots
\\
\bBB 
&UM 
&10 
&1.2
&1
&$10^3$
&\dots
\\
\cCC
&UM 
&10 
&1.5
&1
&$10^3$
&\dots
\\
\dDD
&UM 
&10 
&2
&1
&$10^3$
&\dots
\\
\eEE
&UM
&10 
&1
&1 
&100
&\dots
\\
\fFF
&UM
&10 
&2
&1 
&$3\xx{6}$
&\dots
\\
\gGG
&UM 
&10
&2
&0.03 
&\dots\tablenotemark{d}\tablenotetext{4}{Only low energy spectra well below the cutoff from the FEB are considered.}
&\dots
\\
\hHH
&UM
&10
&2
&1
&\dots\tablenotemark{d}
&\dots
\\
\iII
&UM 
&10
&2
&30 
&\dots\tablenotemark{d}
&\dots
\\
\jJJ
&NL
&10
&1
&1 
&25
&0.08
\\
\kKK
&NL
&10
&1.2
&1 
&250
&0.12
\\
\lLL
&NL
&10
&2
&1 
&$2\xx{6}$
&0.15
\\
\mMM
&NL
&10
&2
&30
&$10^5$
&0.1
\\
\nNN
&NL
&1.002
&1
&1
&$2.2\xx{4}$
&0.8
\\
\oOO
&NL
&1.002
&1.5
&0.01
&$2.8\xx{6}$
&0.8
\\
\tableline
\end{tabular}
\end{center}
\end{table}


\end{document}